\documentclass[a4paper,12pt,oneside]{article}
\usepackage[top=3cm, bottom=3cm, left=3cm, right=3cm,dvipdfm]{geometry}
\usepackage{amsmath,amssymb,graphicx}
\usepackage[dvipdfm]{hyperref}
\usepackage{xcolor}
\hypersetup{
    colorlinks,
    linkcolor={blue!50!black},
    citecolor={red!40!black},
    urlcolor={green!60!black}
}

\usepackage{amsmath,amssymb,graphicx,epsfig,dsfont,color,amsthm}
\usepackage{enumerate}

\def\coex{\global\advance\count1by1}\count1=0

\def\rho{\varrho}
\def\qed{\nobreak\hfill $\Box$\bigskip}

\newtheorem{thm}{\bfseries Theorem}
\newtheorem{claim}[thm]{\bfseries Claim}        
\newtheorem{problem}{Problem}
\newtheorem{opprob}{Open Problem}
\newtheorem{cond}{Condition}
\newtheorem{coroll}[thm]{Corollary}

\newcommand{\prf}{\noindent {\bf Proof. }}

\newcommand{\kiemel}{\textbf}

\newcommand{\N}{\ensuremath{\mathbb{N}}}
\newcommand{\oG}{\ensuremath{\vec{G}}}
\newcommand{\igaz}{{\sc True}\ }
\newcommand{\hamis}{{\sc False}\ }

\begin{document}

\title{Acyclic orientations with degree constraints}

\author{Zolt\'an Kir\'aly\thanks{Research is supported by a grant (no.\ K
    109240) from the National Development Agency
    of Hungary, based on a source from the Research and Technology Innovation
    Fund.
  }\\[-0.8ex]
\small E\"otv\"os Lor\'and University \\[-0.8ex]
\small Department of Computer Science \\[-0.8ex]
\small P\'azm\'any P\'eter s\'et\'any 1/C \\[-0.8ex]
\small Budapest, Hungary, H-1117\\[-0.8ex]
\small \texttt{kiraly@cs.elte.hu}
  \and
  D\"om\"ot\"or P\'alv\"olgyi\thanks{Research is supported by the Lend\"ulet program of the Hungarian Academy of Sciences (MTA), under grant number LP2017-19/2017.
}\\[-0.8ex]
  \small MTA-ELTE Lend\"ulet Combinatorial Geometry Research Group\\[-0.8ex]
	\small E\"otv\"os Lor\'and University \\[-0.8ex]
\small Department of Computer Science \\[-0.8ex]
\small P\'azm\'any P\'eter s\'et\'any 1/C \\[-0.8ex]
\small Budapest, Hungary, H-1117\\[-0.8ex]
\small \texttt{dom@cs.elte.hu}
}

\maketitle

\begin{abstract}
  In this note we study the complexity of some generalizations of the notion of $st$-numbering.
	Suppose that given some functions $f$ and $g$, we want to order the vertices of a graph such that every vertex $v$ is preceded by at least $f(v)$ of its neighbors and succeeded by at least $g(v)$ of its neighbors.
	We prove that this problem is solvable in polynomial time if $fg\equiv 0$, but it becomes NP-complete for $f\equiv g \equiv 2$.
	This answers a question of the first author posed in 2009.
\end{abstract}

\section{Introduction}

In this paper $G=(V,E)$ always denotes an undirected connected graph,
parallel edges are allowed but loops are not.
We use $n=|V|$, $d(v)$ for the degree of a vertex $v\in V$ and
$d(v,Y)$ for the number of edges going from $v$ to $Y$.
For a function $f: V\to \N$, we use the notation $f(X)=\sum_{x\in X}
f(x)$ for any $X\subseteq V$. 

In a digraph, the indegree (number of incoming arcs) of a 
vertex $v$ (or of a set $X\subseteq V$)
is denoted by $\rho(v)$ (or $\rho(X)$, resp.)
and the outdegree is denoted by $\delta(v)$ (or $\delta(X)$, resp.).

The first named author proposed to study the complexity of the following problem in
2009 in Category ``Orientations'' of EgresOpen \cite{egresopen}.

\begin{problem}[\cite{egresopen}]\label{pr1}
Given $G=(V,E)$ and $s,t\in V$ and positive integers $k,\ell$,
decide whether $G$ has an \emph{acyclic} orientation where for every
vertex $v\in V\setminus\{s,t\}$, there are $k$
pairwise arc-disjoint directed paths from $s$ to $v$, and $\ell$
pairwise arc-disjoint directed paths from $v$ to $t$.
\end{problem}

In this paper we settle the complexity of this problem by show that it is NP-complete, already for $k=\ell=2$.

A more general problem that plays a central role in this note, is the following.
We are given $G=(V,E)$ and two functions
  $f: V\to \N$ and $g: V\to \N$ with 
$f(v)+g(v)\le d(v)$ for each $v\in V$. An orientation of $G$ is called
$(f,g)$-\emph{bounded} if $f(v)\le\rho(v)\le d(v)-g(v)$ for each $v\in V$.

\begin{problem}[Degree constrained acyclic orientation problem]\label{dcaop}
  Given $G,f,g$, decide whether $G$ has an \emph{acyclic}
  $(f,g)$-bounded orientation.
\end{problem}

Note that $f(v)$ is a lower bound for the indegree of $v$ while
$g(v)$ is a lower bound for the outdegree of $v$ as we are dealing with
orientations, so $\delta(v)=d(v)-\rho(v)$. 

\begin{claim}\label{claim}
  Problem \ref{pr1} is equivalent to the degree constrained acyclic orientation
  problem if $f(v)=k$ and  $g(v)=\ell$ for all $v\in V\setminus\{s,t\}$;
  and $f(s)=g(t)=0, \; g(s)=f(t)=d(s)$.
\end{claim}

\prf If the prescribed pairwise arc-disjoint directed paths exist,
then obviously
for every $v\in V\setminus\{s,t\}$ we have $\rho(v)\ge k$ and
$\delta(v)\ge \ell$.

Suppose we have an orientation $\oG$ where all $v\in V\setminus\{s,t\}$ have
$\rho(v)\ge k$ and
$\delta(v)\ge \ell$. This orientation defines a (not necessarily unique)
topological order, $V=\{v_1,\ldots,v_n\}$, such that for every directed edge
$v_iv_j$ we have $i<j$ (we may suppose $v_1=s$ and $v_n=t$).
By symmetry and by Menger's theorem, it is enough to prove that for any
$X\subseteq V$ if $s\not\in X$, then $\rho (X)\ge k$. Let $v_i$ be the first
vertex of $X$.
As $s\not\in X$ we have $i>1$.
Clearly $\rho(X)\ge\rho(v_i)\ge k$.\qed

\section{Polynomially solvable cases}

First we examine the special case of Problem \ref{dcaop} when $g(v)=0$ for all $v$, i.e., only lower
bounds on the indegrees are given.
For a set $X\subseteq V$, we call a vertex $x\in X$ a \emph{potential source} 
if $f(x)\le d(x,V\setminus X)$.

\begin{cond}\label{cond}
  For every non-empty set $X\subseteq V$, there exists an $x\in X$ which is a
  potential source for $X$.
\end{cond}

\begin{thm}[Folklore]\label{thm_poli_upper}
There exists an acyclic orientation of $G$ where $\rho(v)\ge f(v)$ for every
$v\in V$ if and only if
Condition \ref{cond} is satisfied. Moreover, such an orientation can be produced
(and simultaneously Condition \ref{cond} can be checked) by a greedy
algorithm. 
\end{thm}

\prf If the orientation exists, then for each $X$ take $x$ as the first vertex
of it (by a topological order). Clearly $f(x)\le\rho(x)\le d(x,V\setminus X)$.

Suppose Condition \ref{cond} is satisfied. 
By applying it to $X=V$, we obtain a vertex $v_1$ with $f(v_1)=0$.
Let $G':=G-v_1$ and for any $v\in V\setminus\{v_1\}$ let $f'(v):=\max(f(v)-d(v_1,v),0)$.
Clearly Condition \ref{cond} is satisfied for the new $G'$ and $f'$, thus, by
induction, there is an acyclic orientation of $G'$ where the indegrees are
lower-bounded by $f'$. The required orientation of $G$ is given by orienting
the edges incident with $v_1$ from $v_1$.

If, at some point of this recursive process, we do not have any
vertex $v$ such that the current $f(v)$ is $0$, then the
set $X$ of the current vertices violates Condition \ref{cond}.
Otherwise we get the orientation required, thus by the first part of the
proof Condition \ref{cond} is satisfied.
\qed

We can go a little bit further.

\begin{thm}\label{thm_upper-or-lower}
  If $G=(V,E)$ and $f(v)g(v)=0$ for every $v\in V$
  (i.e., every vertex has either only a
  lower bound or only an upper bound on the indegree), then we can decide in
  polynomial time whether there is an $(f,g)$-bounded acyclic orientation of
  $G$.
\end{thm}
\prf Let $A=\{v\in V \;|\; f(v)=0, \; g(v)>0\}, \;
B=\{v\in V \;|\; f(v)=0, \; g(v)=0\}, \;
C=\{v\in V \;|\; f(v)>0, \; g(v)=0\}$, by the condition of the theorem these
three sets partition $V$. Call an acyclic orientation ABC if
there is a topological order where the vertices of $A$ come first, then the
vertices of $B$, and finally the vertices of $C$. 
It is easy to observe that if an $(f,g)$-bounded acyclic orientation of
  $G$ exists, then there is another one which is ABC.
Let $G'=G[A]$, and $f'(a)=0, \; g'(a)=\max(g(a)-d(a,V\setminus A),0)$ for $a\in A$.
Furthermore let $G''=G[C]$ and $f''(c)=\max(f(c)-d(c,V\setminus C),0), \; g''(c)=0$
for $c\in C$.
By Theorem \ref{thm_poli_upper}, we can check in polynomial time whether
there is an $(f',g')$-bounded acyclic orientation of
  $G'$ and  whether
there is an $(f'',g'')$-bounded acyclic orientation of
  $G''$.
\qed

We call a vertex $v$ \emph{strict} if $f(v)+g(v)=d(v)$.
The next special case we study is, when every vertex is strict.

\begin{thm}\label{thm_poli_strict}
There exists an acyclic orientation of $G$ where $\rho(v)= f(v)$ for every
$v\in V$ if and only if $f(V)=|E|$ and 
Condition \ref{cond} is satisfied. Moreover, such an orientation can be produced
(and simultaneously Condition \ref{cond} can be checked) by a greedy
algorithm. 
\end{thm}

\prf Observe that if $f(V)=|E|$, then $\rho(v)\ge f(v)$ for every $v$ if
and only if $\rho(v)=f(v)=d(v)-g(v)$ for every $v$.
\qed

We have one more sporadic example where the problem is known to be
in P, see Theorem \ref{k_is_1} in the last subsection.

\section{NP-complete cases}

\begin{thm}\label{NPC}
  The degree constrained acyclic orientation problem
(Problem \ref{dcaop})
  is NP-complete.
\end{thm}

We prove a much stronger theorem about a very restricted version of
Problem \ref{dcaop}. 

\begin{thm}\label{main-NPC}
  The degree constrained acyclic orientation problem is NP-complete even if 
  every vertex $v$ is $\rho$-lower-bounded by 0 or 1
  (i.e., $f(v)\le 1$ and $g(v)=0$)
  except one vertex y, which is strict (i.e., $f(y)+g(y)=d(y)$).
\end{thm}

\prf As the problem is obviously is in class NP, it is enough to show its
hardness. 
We reduce the well-known NP-complete problem {\sc Vertex-Cover} \cite{GJ}
to this restricted version. In the problem {\sc Vertex-Cover}, a graph $G=(V,E)$
and an integer $k\le |V|$ is given and we have to decide whether there is a set
$T\subseteq V$ with $|T|=k$ such that every edge has at least one endvertex
in $T$.

Given $G=(V,E)$ and $k$, we need to construct $G'=(V',E')$ and functions
$f,g$. Let $V'=V\cup E\cup \{y\}$ (where $y\not\in V\cup E$).
Let $m=|E|,\; M=m+1$ and $E'=\{ve \;|\; v\in V,\; e\in E,\; v$ is incident to
$e\} \cup \{ye  \;|\; e\in E\} \cup M\{yv  \;|\; v\in V\}$,
where the last edge-set is meant to have $M$ parallel edges between $y$ and
any vertex $v\in V$. The construction of $G'$ is finished, its degree-function
is denoted by $d'$.

Let $f(v)=0$ for $v\in V$, $f(e)=1$ for $e\in E$ and $f(y)=m+kM$.
Let $g(v)=0$ for $v\in V$, $g(e)=0$ for $ e\in E$ and $g(y)=(n-k)M=d'(y)-f(y)$.
We need to prove that $G$ has a cover $T$ of size $k$ if and
only if $G'$ has an acyclic orientation satisfying these degree bounds.
First suppose $T$ is such a cover and let $E_2\subseteq E$ denote the edges 
with both endvertices in $T$. As $T$ is a cover, every edge in
$E_1=E\setminus E_2$ connects $T$ to $V\setminus T$. Define an order on $V'$ as follows.
We start by putting vertices of $T$ (in any order), then elements of $E_1$, then
$y$, and finally vertices of $V\setminus T$ (in any order). 
We still need to place each $uv\in E_2$,
we put such an edge between $u$ and $v$ (thus it will precede exactly one of
its endvertices). This order defines an acyclic orientation of $G'$
(edges oriented from earlier vertex to the later one). The indegree of any
$e\in E$ is exactly one and $\rho(y)=m+kM$, thus this acyclic orientation is
indeed $(f,g)$-bounded.

For the other direction, suppose there exists an acyclic $(f,g)$-bounded
orientation of $G'$, and fix any topological order.  First we claim that $y$
is preceded by every $e\in E$ and by exactly $k$ elements of $V$ (call this
latter subset $T$). If $y$ is preceded by at most $k-1$ elements of $V$, then
$\rho(y)\le (k-1)M+m<f(y)=kM+m$.  If $y$ is preceded by at least $k+1$
elements of $V$, then $\rho(y)\ge (k+1)M>d'(y)-g(y)=kM+m$. Accordingly,
exactly $k$ vertices of $V$ precedes $y$, and as its indegree is exactly $kM+m$,
necessarily every $e\in E$ must also precede it.

It remains to prove that $T$ is a cover in $G$. Suppose this is not the case,
there is an $e=uv\in E$ with $u\not\in T$ and $v\not\in T$. As $e$ precedes
$y$ and $y$ precedes both $u$ and $v$, $\rho(e)=0<f(e)=1$, a contradiction.
\qed

The graph $G'$ used in the proof is not simple.
However, one can split every edge $e$ of $G'$ with a new vertex $w_e$ and
define $f(w_e)=g(w_e)=1$. This gives the following corollary.

\begin{coroll}\label{simple}
  The degree constrained acyclic orientation problem for \emph{simple} graphs
  is NP-complete even if for
  every vertex $v$ either $f(v)+g(v)=d(v)$ (i.e., $v$ is strict) or
  $f(v)\le 1$ and $g(v)=0$ ($v$ is lower-bonded by 0 or 1).
\end{coroll}

\begin{thm}\label{egres-NPC}
Problem \ref{pr1} is NP-complete.
\end{thm}

\prf We reduce Problem \ref{dcaop} to Problem \ref{pr1}.
Given $G,f,g$, we construct an instance of Problem \ref{pr1} as follows.
First we fix $k=\ell$ as the maximum degree in $G$.
We add two new vertices, $s$ and $t$. Then for each vertex $v$ we add
$k-f(v)$ parallel edges between $s$ and $v$, and we also add $k-g(v)$ 
parallel edges between $t$ and $v$. Let $G'=(V',E')$ denote the resulting
graph (where $V'=V\cup\{s,t\}$).

By Claim \ref{claim}, it is enough to show that the following are equivalent.
\begin{itemize}
\item $G'$ has an acyclic orientation with source $s$ and sink $t$ where for
  every $v\in V$, we have $\rho'(v)\ge k$ and $\delta'(v)\ge k$. 
\item $G$ has an acyclic $(f,g)$-bounded orientation.
\end{itemize}

First suppose we have the above acyclic orientation of $G'$.
After deleting $s$ and $t$, we get an acyclic orientation of $G$ where, for
every $v\in V$, we have $\rho(v)=\rho'(v)-k+f(v)\ge f(v)$, and
$\delta(v)=\delta'(v)-k+g(v)\ge g(v)$.

Next suppose  we have an acyclic orientation of $G$ where for
every $v\in V$ we have $f(v)\le\rho(v)\le d(v)-g(v)$.
To orient $G'$, we keep the
orientation of the original edges and orient the edges of form $sv$ from $s$
to $v$, and the edges of form $vt$ from $v$ to $t$;
resulting in an acyclic orientation of $G'$.
As $\rho'(v)=\rho(v)+k-f(v)$, we get $\rho'(v)\ge k$. We also have
$\delta'(v)=\delta(v)+k-g(v)\ge k$.
\qed

\subsection{Problem \ref{pr1} for small $k$ values}

We proved that Problem \ref{pr1} is NP-complete if $k$ and $\ell$ are
part of the input.
What can we say about its status if $k$ and $\ell$ are fixed  small numbers?

The first case is well known, it was solved in \cite{LEC} where
$st$-numbering was introduced.

\begin{thm}[\cite{LEC}]\label{k_is_1}
  If $k=\ell$ is fixed to one,
  then we can answer Problem \ref{pr1} in polynomial
  time, namely there is a required orientation if and only if $G+st$ is
  biconnected.
\end{thm}

On the other hand, we can prove the following.

\begin{thm}\label{thm_NPC_k=3}
Problem \ref{pr1} is NP-complete for $k=\ell=2$.
\end{thm}

\prf
The problem is obviously in NP.
We will reduce the problem {\sc Not-all-Equal}
3SAT \cite{GJ} to it.
In the problem {\sc Not-all-Equal} 3SAT there are variables $x_1,\dots,x_n$ and
clauses $C_1,\dots,C_m$, each clause contains exactly three literals
(a literal is a variable or a negated variable), and we
need to decide whether there is a truth assignment to the variables such that
every clause has at least one \igaz and at least one \hamis literal in it.
Given a {\sc Not-all-Equal} 3SAT instance, we first construct the multigraph $G$ as follows.

\noindent
$V:=\{s\!=\!a_0, a_1, a_2,\dots, a_{4n+2m}, a_{4n+2m+1}\!=\!t,\; y_1,\dots,y_n, z_1,\dots,z_n, C_1,\dots,C_m\}\cup$
$\cup\{x_1,\dots,x_n,\bar{x}_1,\dots,\bar{x}_n\}$.

First we add edges to make the ``skeleton''.
We add two parallel edges between $a_{2i}$ and $a_{2i+1}$ for
$i=0,\dots,2n+m$. We also add one edge between $a_{2i-1}$ and $a_{2i}$ for
$i=1,\dots,2n+m$.
Then we connect $a_{2i-1}$ to $y_i$ and $y_i$ to $a_{2i}$,
and we also connect $a_{2n+2m+2i-1}$ to $z_i$ and $z_i$ to $a_{2n+2m+2i}$ for
$i=1,\dots,n$.
To finish the skeleton we connect $a_{2n+2j-1}$ to $C_j$ and $C_j$ to
$a_{2n+2j}$ for $j=1,\dots,m$.

\begin{figure}[!ht]
      \centering
      \smallskip
      \includegraphics{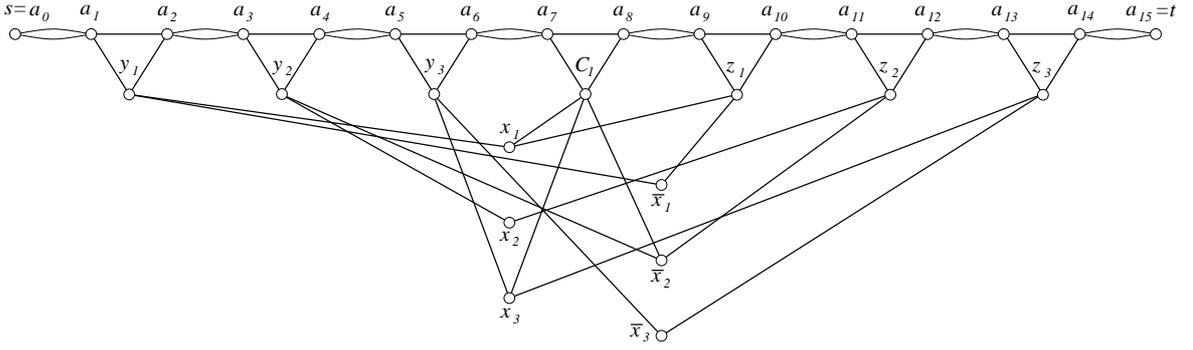}
\begin{center}
      \caption{The construction for $n=3,\; m=1,\; C_1=x_1\vee \bar{x}_2\vee
        x_3$
      (except the parallel edges between the literals and $s$ and $t$ added at the end).}\label{fig_constr}
\end{center}
\end{figure}

We are left to connect the literals to the skeleton.
For $i=1,\dots,n$, we connect both $x_i$ and $\bar{x}_i$ to $y_i$ and $z_i$.
Then for $j=1,\dots,m$, we connect $C_j$ to the three literals it contains.
Finally we add two parallel edges between $s$ and any literal, and also
between $t$ and every literal.

The construction of $G$ is finished, see Figure \ref{fig_constr}.

Note
that the vertices $a_1,\dots,a_{4n+2m}$ have degree four
as well as vertices $y_1,\dots,y_n$ and $z_1,\dots,z_n$, while vertices
$C_1,\dots,C_m$ have degree five.

First we show that if we have an assignment to the variables such that every
clause has at least one and at most two \igaz literals in it, then we can make 
an acyclic orientation of $G$ 
where for every vertex $v\in V\setminus\{s,t\}$ we have $\rho(v)\ge 2$ and
$\delta(v)\ge 2$.
We give the orientation by defining the topological order.
First we define the order of the skeleton:
$$a_0,a_1,y_1,a_2,a_3,y_2,a_4,\dots,y_n,a_{2n},a_{2n+1},
C_1,a_{2n+2},a_{2n+3},\dots,a_{2n+2m-1},C_m,$$
$$a_{2n+2m},a_{2n+2m+1},z_1,a_{2n+2m+2},\dots,a_{4n+2m-1},z_n,a_{4n+2m},a_{4n+2m+1}.$$
Note that the vertices $a_i$ are ordered by the indices, and every
$y_i, z_i, C_j$ is between its two neighbors.
So far each $a_i$ has indegree 2 and outdegree 2 for $1\le i\le 4n+2m$ and
vertices $y_i, z_i, C_j$ have indegree 1 and outdegree 1.
Finally we place all the \igaz literals between $a_1$ and $y_1$ (in any order),
and we place all the \hamis literals between $z_{n}$ and $a_{4n+2m}$
(in any order). First note that due to the parallel edges added at the end,
every literal  has at least two incoming and at least two outgoing arcs.
As any vertex $y_i$ or $z_i$ is preceded by the \igaz literal, and succeeded by the \hamis
literals, every $y_i$ and $z_i$ has
indegree and outdegree exactly two.
Finally, a clause is preceded by its \igaz literals, we have one or two of them,
consequently either its indegree is two and its outdegree is three,
or vice versa.

It remains to show that if the required acyclic orientation exists, then we
have an assignment to the variables such that each clause gets one or two
\igaz values. Take the topological order of a good acyclic orientation.
First we claim that the skeleton is in the same order as we defined in the
previous part. Suppose for a contradiction that $i$ is the smallest index $(1\le i\le 4n+2m-1)$
such that $a_{i+1}$ precedes $a_i$. In this case $a_i$ has at least three
incoming arcs, so this case is impossible. Next suppose that $w$ is the first
vertex from $y_1,\dots, y_n, C_1,\dots, C_m, z_1,\dots, z_n$ which is not
between its two $a_i$-neighbors. If it is before that place, then its
lower-indexed  $a_i$-neighbor has indegree three, if it is after that place,
then its higher-indexed  $a_i$-neighbor has outdegree three. Next observe
that for all $1\le i\le n$, one of $x_i$ and  $\bar{x}_i$  must be before
$y_i$ (consequently before $a_{2n}$) and the other must be after $z_i$
 (consequently after $a_{2n+2m}$), since otherwise either $y_i$ or $z_i$ does
not have the prescribed indegree two.
Now we can define the assignment: a literal is \igaz if it precedes $a_{2n}$.
To finish the proof, observe that if $C_j$ has for example three \igaz
literals, then its indegree is four, so its outdegree is only one.
\qed

\bigskip

The graph $G$ used in the above proof is not simple, actually the answer is
always NO for a simple graph, as the second vertex cannot have two incoming
arcs.
However, one can split any edge $e=uv$ of $G$ with a new degree four vertex $w_e$ that is connected to $s,u,v,t$.
Such a $w_e$ must necessarily be between $u$ and $v$, thus their in- and outdegrees will not be affected.
This way we can obtain a graph that is almost simple - only edges adjacent to $s$ or $t$ might have multiplicity $2$.

By splitting $s$ into two vertices, $s_1$ and $s_2$, and similarly $t$ into $t_1$ and $t_2$, and dividing the multiple edges among them, we obtain that the following problem is NP-complete.

\begin{problem}
	Given a simple graph $G=(V,E)$ and $s_1,s_2,t_1,t_2\in V$,
	decide whether there is an order of the vetrtices such that
        every vertex $v\in V\setminus\{s_1,s_2,t_1,t_2\}$ has at least
        $2$ neighbors preceding it and also at least $2$ neighbors succeedenig
        it.
\end{problem}

Some of the questions left open are the following.

\begin{opprob}
Is Problem \ref{pr1} in P for the special case $k=2$ and $\ell=1$?
\end{opprob}

\begin{opprob}
  Is Problem \ref{dcaop} in P for the special case $f(v)+g(v)\le 2$ for all
  $v\in V$?
\end{opprob}

\begin{opprob}
  Is Problem \ref{dcaop} in P for the special case $f(v)+g(v)\le 3$ for all
  $v\in V$?
\end{opprob}

\begin{opprob}
	Is Problem \ref{dcaop} in P for the special case $f(v)g(v)\le 1$ for all
	$v\in V$?
\end{opprob}

\end{document}